\documentclass[aps,prb,twocolumn]{revtex4}

\usepackage{amsmath}
\usepackage{amssymb}
\usepackage{graphicx}

\begin{document}
\newcommand{\bstfile}{osa} %alternative styles: osa, prasty or revtex
\newcommand{\bibs}{D:/JCP_paper/604831JCP_archive}
\title{Mechanisms of reversible photodegradation in Disperse Orange 11 dye doped in PMMA polymer}

\author{Natnael B. Embaye}
\email{nembaye@wsu.edu}

\author{Shiva K. Ramini}
\email{rshiva@wsu.edu}

\author{Mark G. Kuzyk}
\email{kuz@wsu.edu}

\affiliation{Department of Physics and Astronomy, Washington State University, Pullman, WA 99164-2814}

\date{\today}

\begin{abstract}
We use amplified spontaneous emission (ASE) and linear absorption spectroscopy to study the mechanisms of reversible photodegradation of 1-amino-2-methylanthraquinone (Disperse Orange 11 - DO11) in solid poly(methyl methacrylate).  Measurements as a function of intensity, concentration, and time suggest that ASE originates in a state (be it a tautomer or vibronic level) that can form a dimer or some other aggregate upon relaxation, which through fluorescence quenching leads to degradation of the ASE signal.  Whatever the degradation route, a high concentration of DO11 are required and the polymer plays a key role in the process of opening a new reversible degradation pathway that is not available at lower concentration or in liquid solution.  We construct an energy level diagram that describes all measured quantities in the decay and recovery process and propose a hypothesis of the nature of the associated states.
\end{abstract}

\maketitle

\section{Introduction}
All materials are damaged by light exposure if the light intensity is high enough or the time long enough, be it the photobleaching process that causes the vibrant colors of clothing to fade or the catastrophic laser damage of optical components in a laser cavity.  As such, research has focused on the development of an understanding of photo-damage and methods to mitigate such damage.\cite{exarh98.01,wood03.01} An approach to making more robust materials is to use our understanding of the mechanisms of photo-damage as a guide for new design paradigms.  Alternatively, it may be possible to process existing materials in new ways to make them better.

Optical damage can be separated into intrinsic and extrinsic causes.  Impurities or defects that act as centers of energy deposition can be removed by purification of the starting materials, keeping them in a clean environment, and varying the processing conditions.  Such extrinsic damage mechanisms in a dye-doped polymer can be reduced by filtering and distilling the liquid monomer used in making the polymer, purifying the dopant dyes, and making the materials in a clean environment under highly controlled processing conditions.\cite{kuzyk06.06}  Intrinsic properties, such as the strength of absorption by the molecules and the polymer can only be reduced by judiciously choosing the polymer host material and the dopant molecules.

This paper focuses on the idea that it may be possible to condition existing materials using light, radiation, and thermal cycling to make them better.  Our work is based on a serendipitous discovery that certain materials self-heal after photodegradation, and that they become more robust in the process.  The mechanisms of the cause of decay and recovery is an interesting scientific question that also has important implication for designing materials that better resist optical damage.

Photodegradation and recovery has been observed in many materials using a variety of measurement techniques.  Peng and coworkers observed fluorescence decay and recovery in rhodamine-doped and pyrromethene-doped polymer optical fibers.\cite{Peng98.01}  The dye-doped fibers fully recovered provided that the degree of degradation was not too high.  Howell and coworkers first demonstrated that 1-amino-2-methylanthraquinone (Disperse Orange 11, or DO11) dye-doped polymer would make a full recovery even when the degree of degradation was nearly 100\%, as measured with amplified spontaneous emission (ASE).\cite{howel02.01} However, no recovery of the same dye in liquid solution was observed at all decay levels. This suggested that the host polymer plays a crucial role in the self-healing process.\cite{howel04.01}  More recently, Zhu and coworkers observed full recovery of the octupolar dye AF455 doped in PMMA polymer, as measured with two-photon fluorescence.\cite{zhu07.01} The fact that such a broad range of dyes show reversible photodegradation suggests that molecules, which normally decay irreversibly, can be made to recover when incorporated into a host polymer.\cite{zhu07.02}

An intriguing property of the DO11/PMMA material is that the photodegradation rate decreases after the material experiences a decay and recovery cycle; and, the ASE efficiency increases.\cite{howel02.01,zhu07.02}  Laser cycling and gamma irradiation has recently been observed to result in even a more robust material, suggesting that synergism between the two processes may be used to optimize a material's resistance to future damage.\cite{kuzyk07.02} This is consistent with the observation that gamma irradiation alone can make electro-optic modulators more reliable.\cite{taylo05.01}

The present work focuses on the first step in the development of a fundamental understanding of the mechanisms responsible for photodegradation and subsequent recovery in the DO11/PMMA system, with the long-term goal of providing guidelines for making new and more robust materials as well improving existing materials.

\section{Experiments}

\begin{figure}[htb]
\centerline{\includegraphics{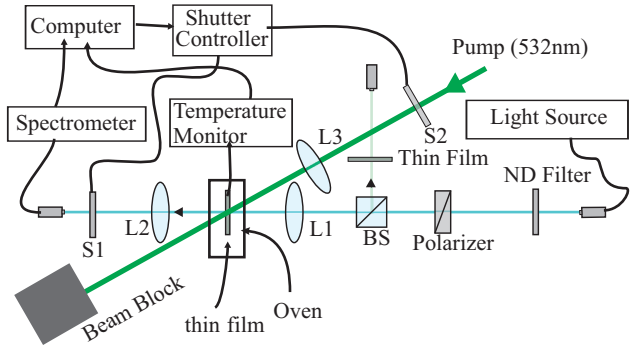}}\caption{
\label{fig:Absorption_setup} The absorbance experiment is computer
controlled and automated.}
\end{figure}

We use amplified spontaneous emission and linear absorption spectroscopy as probes of the dye doped polymer.  Due to the high optical density of DO11/PMMA, the linear absorption experiments use thin-film samples that are made by squeezing a bulk sample to micron thickness on a transparent substrate. Figure \ref{fig:Absorption_setup} shows a diagram of the experiment.

An Ocean Optics miniature deuterium tungsten halogen light source with a spectral range of 200nm-1100nm was used to probe the spectrum.  It is sent through a neutral density filter to adjust the intensity.  After passing the polarizer, a portion of the beam is sent to a detector that monitors the integrated power of the light source.  Two lenses are used to collimate the beam.  The sample is mounted in an oven so that its temperature can be controlled.  A shutter (S1) is used to block the Ocean Optics fiber spectrometer when the pump light is turned on.  The pump beam at 532nm (Frequency doubled Nd:YAG laser, 5ns pulses) is focused in the sample where it spatially overlaps with the white light source.  The pump light can be blocked with a shutter (S2).

First, the transmitted white light spectrum is measured through a clean substrate.  Subsequently, the transmitted spectrum is measured through a thin-film sample coated on an identical substrate.  The ratio of the two spectra is used to determine the absorption spectrum.  The experiment is automated under computer control.

With the Shutter S1 closed, shutter S2 is opened and the sample pumped for a fixed time interval.  Shutter S2 is closed and Shutter S1 is opened long enough to measure an absorption spectrum, after which shutter S1 is closed and S2 opened for the next cycle of exposure.  The polarizer can be rotated to determine the absorption spectrum along and perpendicular to the polarization of the pump beam.

\begin{figure}[htb]
\centerline{\includegraphics{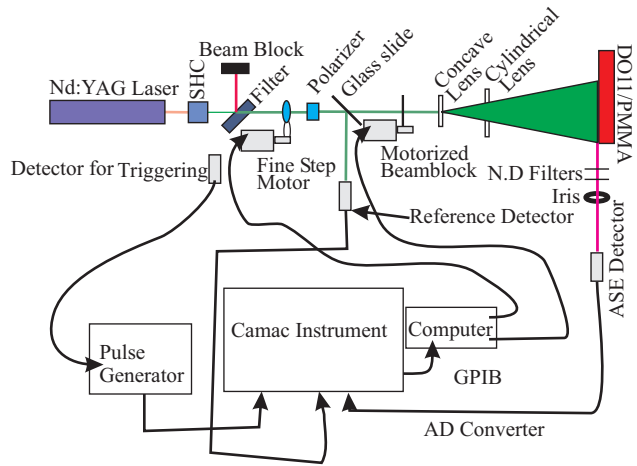}}\caption{
\label{fig:ASE_Setup} The ASE experiment. All the detectors, the
pulse generator, the Camac instrument and the motorized half-wave
plate are computer controlled for automation.}
\end{figure}
Figure \ref{fig:ASE_Setup} shows the ASE experiment.  A reflectance filter that removes the fundamental wavelength is placed immediately downstream of the Second Harmonic Crystal (SHC).  The polarized second harmonic beam (532nm) passes through a half wave plate whose orientation is controlled with a fine stepper motor, thus allowing for automated control of the polarization angle.  The intensity leaving the polarizer is thus a function of the half-wave plate's orientation.  A glass slide picks off a small portion of the beam, which is read by the reference detector and used by the computer to determine the adjustment angle to compensate for long-term drift of the laser power.  This feedback loop is initially adjusted to pass about 80\% of the laser power, thus being able to compensate for drift on the order of $\pm$20\%.

The beam then passes a motorized beam block, is de-focused by a concave lens and focused by a cylindrical lens to a line on the surface of a cylindrical dye-doped sample.  The sample is made by polymerizing a mixture of liquid MMA monomer with dissolved DO11 chromophores in a test-tube, which when shattered, yields a cylindrical sample with an optically smooth surface.  The two ends are cut and polished to make them optically smooth.  The ASE light that leaves the sample through one of the polished ends is measured with a detector and the signal recorded by computer.  The sample can be placed in an oven (not shown) to do temperature-dependent studies.

\section{Results and Discussion}

\begin{figure}[htb]
\centerline{\includegraphics{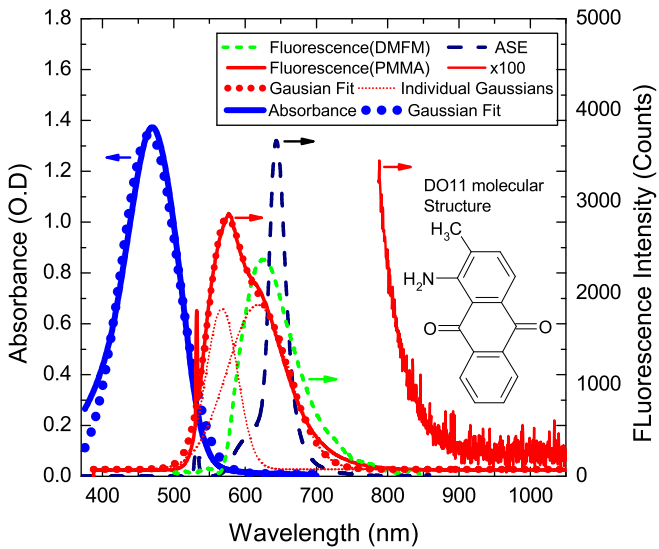}}\caption{
\label{fig:abs_flu_ASE} Absorption, fluorescence (pumped with
20$\mu$J/pulse) in PMMA and in solution and ASE spectrum (pumped
with 2$m$J/pulse) of DO11/PMMA at 9g/l concentration.}
\end{figure}
The left-most peak in Figure \ref{fig:abs_flu_ASE} shows the
absorption spectrum of the DO11 chromophore in PMMA polymer and
the points represent a single Gaussian peak fit.  The inset shows
the molecular structure.  At low pump power, the measured
fluorescence spectrum in PMMA polymer is broad and is not the
mirror image of the absorption spectrum as is common.  (The spike
at 532nm is scattered light from the pump.)  The two thin-dotted
curves are the individual peaks in a two-Gaussian peak fit and the
points coincident with the fluorescence spectrum are the sum of
the two individual peaks.  The short-dashed curve is the
fluorescence spectrum of DO11 in dimethylformamide(DMFM) solvent
and the long-dashed peak is the ASE spectrum of DO11 in PMMA
polymer.  The far right spectrum is the tail of the fluorescence
spectrum of DO11/PMMA magnified 100 times.  No other peaks are
observed in the fluorescence spectrum up through 1050nm, so if
there are any tautomer peaks in this region, they are well below
the resolution of our spectrometer.  Note that the fluorescence
spectrum of DO11 in dimethylsulfoxide (DMSO) and butyrolactone is
observed to be similar to the fluorescence spectrum of DO11 in
DMFM, but only the fluorescence spectrum of DO11 in chlorobenzene
shows the same shoulder structure as DO11/PMMA.

Anthraquinone molecules with structures similar to DO11 are known to photo-tautomerise, where light excitation causes one of the protons on the $H_2N$ group to jump to the adjacent oxygen forming an $OH$ and leaving behind  an $NH$. A large separation between the peak absorbance and fluorescence is indicative of photo-tautomerization.  For example, in 1-(acylamino)anthraquinones, the absorption and fluorescence peak is separated by about 100nm and the tautomer fluorescence peak is more than 200nm from the absorption peak.\cite{smith91.01} Thus, the fluorescence spectrum has two overlapping peaks whose relative strengths can be changed by the choice of solvent.\cite{smith91.01}

These overlapping peaks are similar in appearance to the peak and shoulder in the fluorescence spectrum of DO11/PMMA except that the two peaks that are found by fitting to two Gaussian functions are found to be separated by about 55nm, two to three times smaller than for 1-(acylamino)anthraquinones.  Interestingly, the fluorescence component peak near 620nm of DO11/PMMA is similar in position and shape to the fluorescence spectrum in solution.

There are several possible explanations for the shoulder in the DO11/PMMA fluorescence spectrum.  First, it could be a vibronic state.  However, this same shoulder does not appear in linear absorption spectrum.  Secondly, the shoulder might originate from a tautomer state. However, the separation between the peaks is not as large as is typical for tautomer states in similar molecules.  For example, Shynkar and coworkers have reported on picosecond time-resolved fluorescence studies of reversible excited-state intramolecular proton transfer in 4'-(Dialkylamino)-3-hydroxyflavones and find that the tautomer and normal peak in the fluorescence spectrum is separated by about 0.28 eV,\cite{shynk03.01} in contrast to our peak separation of 0.18eV.  Third, perhaps the molecules are forming aggregates that distort the fluorescence spectrum.  However, it is peculiar that the same shoulder is not found in the linear absorption spectrum.

All of the obvious explanations of the shapes of the spectra are problematic.  Some of these issues might be resolved by the fact that the molecules are embedded in a host polymer, and that this may complicate the problem.  It is well known that polymers are microscopically inhomogeneous leading to a distribution of sites so that each molecule that is embedded in a polymer sees a different local environment.\cite{ghebr93.01,ghebr95.04} In contrast to a liquid, where broadening of the energy levels are due to stochastic collisional processes, in a polymer, broadening can be dominated by a distribution in local fields due to the distribution sites.

Another complication arises in sample preparation.  When making thin films through spin coating, the polymer and dyes are dissolved in a solvent to make a thick viscous solution that forms a thin film when placed on a spinning substrate once the solvent evaporates.\cite{poga94.01,poga95.01} Such materials have been extensively studied and well understood.  However, in our studies, we use large bulk samples, so solvents are problematic because they are not easy to remove from the bulk.  For these reasons, our samples are made by polymerizing a mixture of liquid MMA monomer and the dyes in the presence of initiator, in a process that was originally developed for making preforms used in drawing polymer fibers.\cite{garve96.02,garve96.01}

During the polymerization process, some types of molecules are
known to be chemically altered.  This is not observed in DO11.  As
the polymer forms, smaller and smaller pockets of MMA are found in
larger and larger amounts of polymer.  Given the difference in
solubility of the dye in monomer and polymer, however small, could
lead to regions of higher dye concentrations.  All of these
complications make it more difficult to interpret the meaning of
the fluorescence data in a polymer matrix compared to the simpler
case of dyes in solution.

For our purposes, the most important feature of the ASE spectrum is that the peak appears to be associated with the lower-energy component of the fluorescence peak in DO11/PMMA.  Since our goal is to determine the mechanisms of self-healing, the details of the light production process is not central to our argument and determining the mechanisms is well beyond the scope of the present studies.  However, understanding the mechanisms of light production clearly warrants further study in the future if the process of degradation is to be fully understood.  Such studies will undoubtedly also add to our understanding of the recovery process, and therefore would be a crucial part of building a broader theory.

Thus, we propose the following hypothesis for the cycle of decay and healing.  ASE originates in either a particular excited state vibronic level or in an excited state tautomer as it makes a transition to the ground state of the molecule or the tautomer.  In both cases, a relaxation process follows the original excitation to populate a relatively long-lived excited state.  The ground state tautomer or vibronic level subsequently relaxes back to the DO11 ground state.  Thus, the absorption peak represents the energy difference between the ground and excited state of the DO11 molecule while the peak fluorescence wavelength yields the energy difference between the ground and excited state tautomer or two vibronic levels.

In our present work, where we use a 5ns pump, the fluorescence spectrum shoulder is at a wavelength of 623nm.  As the pump intensity is increased, an ASE peak forms to the red of the shoulder at 645nm, which is a transmission window in most polymers because it falls between the overtones of the CH stretch modes.  In the 25ps pump measurements reported by Howell and coworkers, the shoulder was observed at 630nm and the ASE peak at 650nm.\cite{howel02.01} The polymer may be partially responsible for absorbing the ASE light generated at other wavelengths.  This material is thus potentially useful for making light sources to be used in local area network applications, which uses 650nm as a standard due to a transparency window here in graded polymer optical fibers.

\begin{figure}[htb]
\centerline{\includegraphics{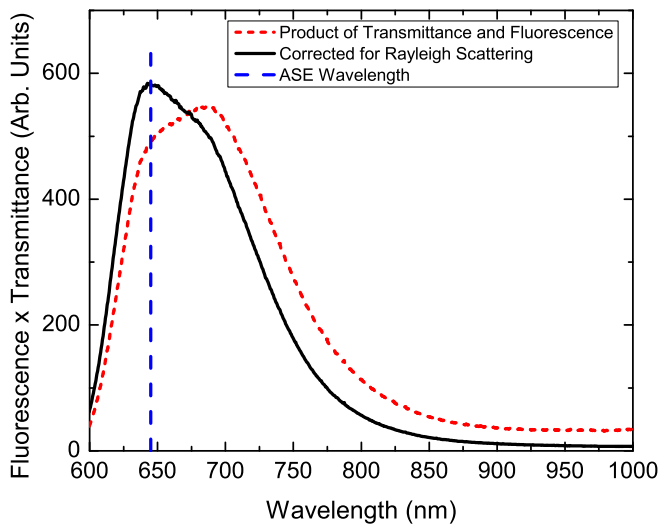}}\caption{
\label{fig:AbsFluor} Product of the fluorescence spectrum and the
transmittance spectrum of a 12mm thick cylinder of PMMA/DO11
(dashed curve) and corrected for Rayleigh scattering (solid
curve).}
\end{figure}
Figure \ref{fig:AbsFluor} shows the product of the transmission spectrum and fluorescence spectrum for a sample that is about 12mm thick.  The sample thickness is comparable to the length of the line focus used for ASE measurements.  Since both Rayleigh scattering and absorption both contribute to attenuation over this long sample, we have corrected the data for the $\lambda^4$ dependence of Rayleigh scattering to isolate the contribution due only to optical absorbtion.  We argue that the peak of the adjusted spectrum is the appropriate one for determining the ASE wavelength due to the fact that a stimulated process is less sensitive to scattering than absorption.  Indeed, the peak of the product function agrees well with the measured ASE wavelength, as shown by the vertical dashed line.

The ASE intensity depends on the polarization of the pump light.  When the pump field is polarized along the line focus, the ASE intensity vanishes.  When the polarization axis is perpendicular to the line focus, the ASE signal peaks.  This is consistent with ASE being generated by dipole radiation, where the intensity is maximal in the plane perpendicular to the dipole axis. As such, the polarization is set perpendicular to the excitation line for all ASE experiments.

\begin{figure}[htb]
\centerline{\includegraphics{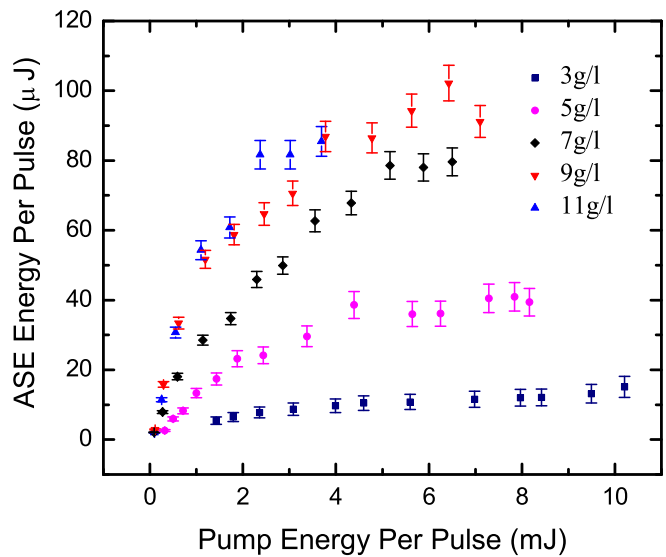}}\caption{
\label{fig:Gaggregate} ASE as a function of pump energy for 3, 5,
7, 9 and 11g/l concentrations.}
\end{figure}
Figure \ref{fig:Gaggregate} shows the ASE energy per pulse as a function of the pump energy per pulse for several different concentrations of DO11.  The ASE signal does not get larger when the DO11 concentration is increased above 9g/l.  At 11g/l, DO11 crystals are visually observed to form during polymerization of the monomer/dye solution.  The fact that the ASE signal levels off at 9g/l DO11 is consistent with the well-known fact that aggregation quenches fluorescence and ASE.

\begin{figure}[htb]
\centerline{\includegraphics{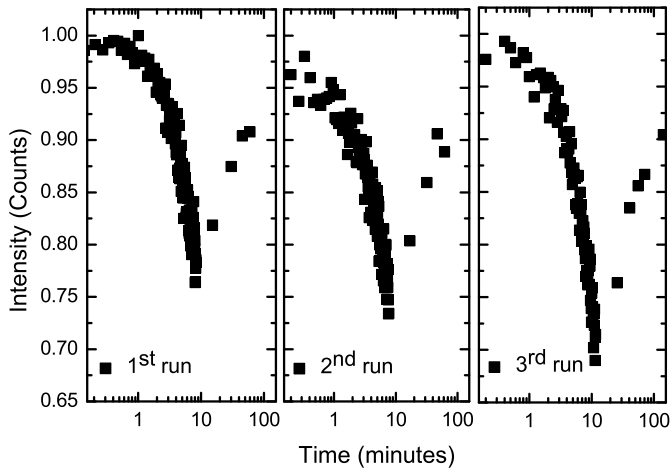}}\caption{
\label{fig:logscale_9g_per_lt} Photodegradation and recovery for a
9g/l sample of DO11/PMMA for three successive cycles of pumping
and rest. The pump energy per pulse is 2mJ. Note that the sample
will fully recover when allowed to heal for a long enough period
of time.}
\end{figure}
Figure \ref{fig:logscale_9g_per_lt} shows decay and recovery of ASE over three successive cycles.  Each successive run is plotted on the same scale as the first run.  It is clear that the sample has fully recovered after the second run from a comparison of the beginning of the third run with the beginning of the first run.  We commonly make the observation that when a DO11/PMMA sample is permitted to rest over a long enough time period, the sample recovers to a level better than at the starting point.  This observation suggests that it may be possible to make materials better with cycling.\cite{howel02.01,zhu07.01} We note that our ASE measurement uses a 5ns pulse width laser, which gives similar results found by Howell using 25ps pulses.\cite{howel02.01}

\begin{figure}[htb]
\centerline{\includegraphics{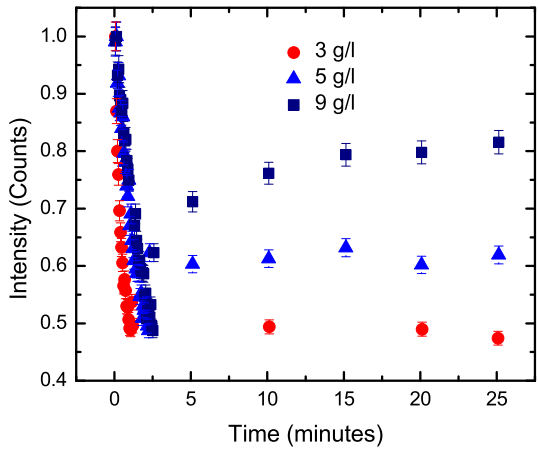}} \caption{
\label{fig:3_5_9g_per_l} ASE as a function of time for 3, 5, and
9g/l concentrations.}
\end{figure}
Figure \ref{fig:3_5_9g_per_l} shows ASE as a function of time for three concentrations.  During degradation, the pump laser fires continuously at 10Hz until the ASE signal reaches half its initial value, at which point the pump laser is blocked to allow the sample to heal.  The ASE intensity is measured every 5 minutes during which time the shutter is opened for 10 seconds to monitor the recovery kinetics without causing appreciable degradation.  The degree of recovery at 25 minutes is strongly correlated with the concentration.  As such, we speculate that  self-healing requires chromophore concentrations near the aggregation threshold.

To summarize, in liquid solutions or at low enough concentrations in PMMA polymer such that the DO11 molecules do not aggregate, photodegradation is observed without subsequent recovery.  This implies that a degradation process that is independent of interactions between DO11 molecules is responsible.  This observation suggestw that at high-enough concentrations where the molecules are known to aggregate, a different degradation process is favored at the expense of the former process.  Note that in Figure \ref{fig:3_5_9g_per_l}, the recovery time constant is the same for each concentration, but the degree of recovery varies.  However, the rate at which the ASE signal decreases gets smaller as the concentration increases because a larger portion of the population is self healing. This is consistent with the notion that the fractional amount of recovery is related to the fraction of molecules whose fluorescence is quenched due to dimer formation.

The model that follows will focus on this dominant mechanism of photodegradation at high concentrations, assuming that the low-concentration degradation process is negligible.

It is well known that in certain proton transfer materials, dimers can form due to excited state double proton transfer after photo-excitation.\cite{ashra75.01,ingha73.01} Since ASE quenching is known to be associated with aggregation, and recovery is observed only at higher concentrations, we propose the hypothesis that the photodegradation mechanism at high concentrations is tied to dimer formation, according to the following scenario.\cite{kuzyk06.06} After absorbing a photon, and emitting ASE light, the molecule finds itself in a ground state (be it a tautomer or a higher energy level of a vibronic state in the ground state manifold).  If the DO11 molecules are close enough together, double proton transfer will occur.  Alternatively, it is also possible that the molecules will be attracted through electrostatic dipolar forces by nearby tautomers, if tautomers are formed and have a larger dipole moment than the DO11 molecule.  Since the DO11 concentration is near the aggregation limit, a large population of molecules have proximate neighbors.  The dimer energy, due to the attractive binding force or double proton exchanges, is lower in energy than the tautomer energy and thus forms a meta-stable state.  If a dimer does not contribute to fluorescence, it is effectively removed from the population of molecules that generate ASE.

We use linear absorption spectroscopy to test the dimer hypothesis.  First, however, we need to eliminate the possibility that orientational hole-burning is responsible.  If the molecules reorient randomly when they de-excite, the polarized excitation light can have the effect of depleting the population of molecules that are originally aligned with the pump laser's electric field.  A net population shift would result in decreased optical absorbance along the pump's polarization and an increase in the perpendicular direction.
\begin{figure}[htb]
\centerline{\includegraphics{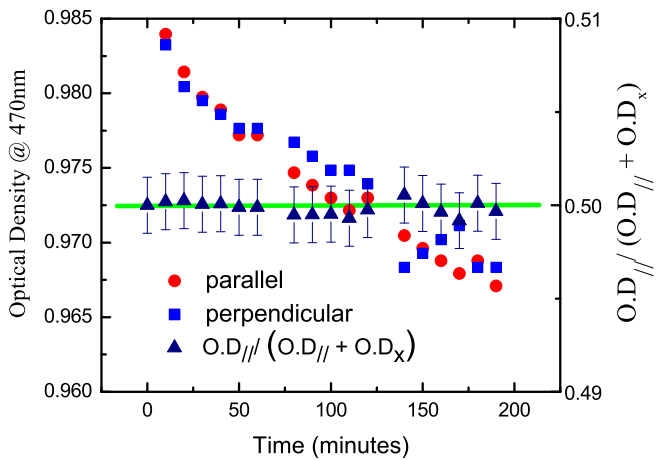}} \caption{
\label{fig:ODratio} Absorbance of DO11/PMMA parallel and
perpendicular to the pump beam polarization and the anisotropy
parameter.}
\end{figure}
Figure \ref{fig:ODratio} shows the optical density, measured with the Experiment shown in Figure \ref{fig:Absorption_setup} at the wavelength of peak absorption, for parallel and perpendicular polarizations.  Also plotted is the polarization ratio, which is a direct measure of the anisotropy.  The anisotropy parameter shows no evidence of orientational hole burning.

A collection of non-interacting identical molecules shows a single absorption peak, as shown in Figure \ref{fig:abs_flu_ASE}.  When two molecules form a dimer, the interaction causes the ground and excited states to split into two states.  When dimers are present, we would thus expect the linear absorbtion spectrum to develop two peaks on either side of the single-molecule peak.  As the population of dimers increases, the main peak would drop due to a decrease in the single molecule population.

\begin{figure}[htb]
\centerline{\includegraphics{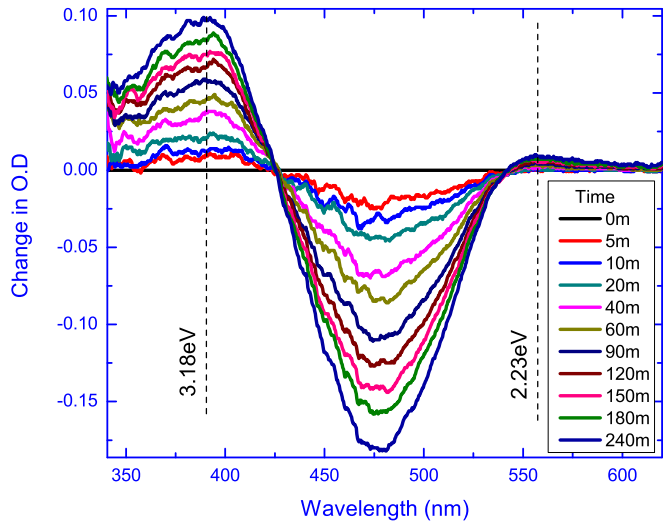}} \caption{
\label{fig:ABSvsTime} Absorbance of DO11/PMMA as a function of
pump laser exposure, with the initial absorption spectrum at $t=0$
subtracted.}
\end{figure}
Since the ground state energies of the individual molecules, tautomers, and dimers might be comparable to thermal energies, all three species could be present even in a pristine sample at room temperature.  So, the best method for detecting a change in populations is to reference all spectra to the original absorption spectrum of a fresh sample.  Figure \ref{fig:ABSvsTime} shows the absorption spectrum as a function of exposure time, with the initial spectrum subtracted.

The depletion of non-aggregated DO11 molecules is clear from the large dip at 470nm (2.64 eV).  The two peaks at either side of the dip at 556nm (2.23 eV) and 390nm (3.18 eV) are consistent with dimer formation.  The two excited state energies of the dimer can be deduced from the positions of the new peaks in the dose-dependent linear absorbance measurements.

\begin{figure}[htb]
\centerline{\includegraphics[width=8cm]{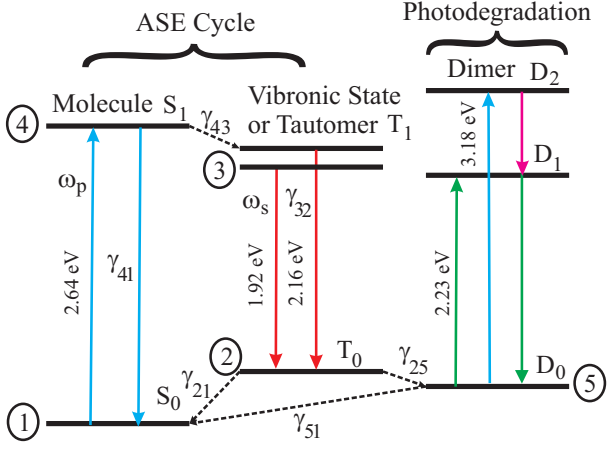}}
\caption{ \label{fig:DO11energylevel} Proposed energy level
diagram of the ASE process, photodegradation and recovery.  The
circled numbers label the important states.}
\end{figure}

Similar measurement have been done as a function of temperature rather than as a function of laser exposure to determine the energy difference between the ground state of the system and the ground state of the depleted species.  The data look similar to Figure \ref{fig:ABSvsTime}, but with slightly shifted peak energies.  According to Figure \ref{fig:DO11energylevel} the temperature dependence of the satellite peak amplitudes relative to the dip magnitude will be related to the population fraction between state $5$ and $1$ according to
\begin{equation}
\frac {N_5} {N_1} = \exp \left[ - \frac {E_5 - E_1} {kT} \right].
\end{equation}
An Arrhenius plot yields an energy difference of about 0.03eV.  This compares favorably with the results of Ingham and El-Bayoumi who got an activation energy of 0.04eV.\cite{ingha73.01}  While the molecular structure studied by Ingham and El-Bayoumi is not the same as DO11, both molecules share the property that proton exchange occurs.  Thus, the similarity between the energetics of our observations and those of similar systems is consistent with our hypothesis of dimer formation in DO11/PMMA, though by no means is this proof.  A more detailed study of the temperature-dependence will be reported in the future.

Figure \ref{fig:DO11energylevel} shows an energy level diagram that schematically represents the photodegradation and self-healing process as we have deduced from dose-dependent linear absorption, ASE and fluorescence.  The ground state DO11 population, $n_1$\footnote{All lower case symbols represent population fractions such that $\sum_i n_i = 1$.}, is depleted by excitation to the excited state (state 4) of population $n_4$ by the pump pulse at an intensity-dependent rate, $\omega_p$, and fed by the tautomer ground state (state 2) of population $n_2$ at a rate $\gamma_{21}$, the dimer ground state (state 5) of population $n_5$ at a rate $\gamma_{51}$, and from decay from the excited state with population $n_4$ at a rate $\gamma_{41}$, leading to the rate equation
\begin{align}\label{eq:n1}
\frac{dn_1(t)}{dt}=-\omega_p  n_1(t) + \gamma_{41}n_4(t) + \gamma_{21}n_2(t) +\gamma_{51}n_5(t).
\end{align}

The population of the tautomer, $n_2$, is depleted by de-excitation to either the molecule ground state at rate $\gamma_{21}$ or forms a dimer of population $n_5$ at a rate $\gamma_{25}$; and, the tautomer population is fed at the rate $\gamma_{32}$ due to spontaneous fluorescence emission from the excited state tautomer, of population $n_3$ (state 3), and from stimulated emission at a rate $\omega_s$.  We note that both ASE and spontaneous emission originates from the same band (shown in Figure \ref{fig:abs_flu_ASE}), but the energy level diagram shows only the peak fluorescence photon energy ($2.16 eV$) and the ASE energy(1.92eV).  The population $n_2$ thus obeys the rate equation,
\begin{align}\label{eq:n2}
\frac{dn_2(t)}{dt}=(\gamma_{32}+\omega_s) n_3(t) - (\gamma_{21} + \gamma_{25})n_2(t).
\end{align}
Note that the ASE intensity is proportional to the excited tautomer population, $n_3$, and the rate at which the ASE intensity drops due to photodegradation is related to rates $\gamma_{25}$ and $\gamma_{21}$.

$\omega_s$, the rate of decay of the excited state tautomer due to stimulated emission is given by,\cite{embay07.01}
\begin{equation}\label{stimulatedEmit}
\omega_s = \gamma_{32} \left( \frac {a} {2L} \right)^2 \left( \exp \left[ n_3(t) N \sigma L \right] - 1 \right) ,
\end{equation}
where $L$ is the length of the excitation line, $a$ the width of the excitation line, $\sigma$, the absorption cross-section, and $N$, the density of dye molecules in the polymer.  Note that we define $n_i$ to be fractional population so that $N n_3$ is the number density of excited state tautomers.

Similarly, we apply the energy level diagram to obtain the rate equations for the three remaining state populations $n_3$, $n_4$ and $n_5$,
\begin{align}\label{eq:n3}
\frac{dn_3(t)}{dt}= - (\gamma_{32}+\omega_s) n_3(t)+ \gamma_{43}n_4(t),
\end{align}
\begin{align}\label{eq:n4}
\frac{dn_4(t)}{dt}=\omega_p n_1(t) - (\gamma_{43}+\gamma_{41})n_4(t),
\end{align}
and
\begin{align}\label{eq:n5}
\frac{dn_5(t)}{dt}=\gamma_{25}n_2(t) - \gamma_{51}n_5(t).
\end{align}
The set of 5 coupled equations can be numerically solved to determine the populations of each state as a function of time.  However, the equations can be greatly simplified under our experimental conditions, as follows.

For fixed pump power, the ASE intensity is proportional to the excited state tautomer population $n_3$, so a change in the ASE power is a measure of a change in $n_3$.  Recalling that the dimer formation rate $\gamma_{25}$ leads to photodegradation by removing population from the ASE cycle, and recovery is governed by the rate at which the dimers break up, $\gamma_{51}$, it is clear from the data (Figure \ref{fig:3_5_9g_per_l}) that the recovery rate (self-healing takes hours) is the longest of all processes.  Similarly, given that all the other time constants are on the order of picoseconds,\cite{ernsr86.01} Figure \ref{fig:3_5_9g_per_l} implies that the degradation rate (50\% degradation takes minutes) is about a couple of orders of magnitude larger than the healing rate, but still much smaller than all other rates.  Thus, the fast processes reach equilibrium quickly, and so we argue that the observed dynamics on our experimental time scales depend only on $1/\gamma_{25}$ and $1/\gamma_{51}$.

When the ASE cycle has reached equilibrium  so that $\dot{n}_1 = \dot{n}_2 = \dot{n}_3 = \dot{n}_4 \approx 0 $; but, when the population of the dimer is still negligible ($n_5 \approx 0$), Equations \ref{eq:n1} to \ref{eq:n5} become:
\begin{align}\label{eq:n12}
\omega_p  n_1(t) = \gamma_{41}n_4(t) + \gamma_{21}n_2(t) ,
\end{align}
\begin{align}\label{eq:n22}
n_2(t) = \frac {\gamma_{32}+\omega_s} {\gamma_{21} + \gamma_{25}} n_3(t),
\end{align}
\begin{align}\label{eq:n32}
n_3(t) = \frac {\gamma_{43}} {\gamma_{32}+\omega_s} n_4(t),
\end{align}
\begin{align}\label{eq:n42}
n_4(t) = \frac {\omega_p} { \gamma_{43}+\gamma_{41}} n_1(t),
\end{align}
and
\begin{align}\label{eq:n52}
\frac{dn_5(t)}{dt}=\gamma_{25}n_2(t) ,
\end{align}
respectively, which can be solved algebraically.

\begin{figure}[htb]
\centerline{\includegraphics{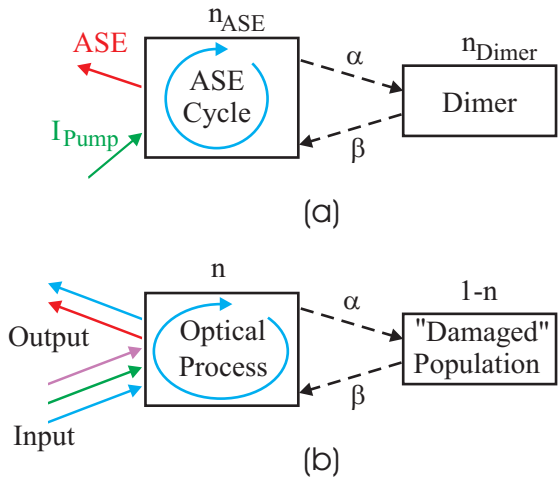}} \caption{
\label{fig:DO11-two-level} A two state model that represents the
photodegradation process.}
\end{figure}

The fraction of the population that is found in the ASE cycle, $n_{ASE}$ is given by
\begin{equation}\label{eq:ASEcycle}
n_{ASE} = n_1 +n_2 + n_3 + n_4,
\end{equation}
with the remaining population, $n_5$, in the dimer state.  On time scales that are long compared with the time it takes the ASE cycle to come to equilibrium but short compared with photodegradation and recovery, the dynamics can be described by a two state model as diagramed in Figure \ref{fig:DO11-two-level}a.

The rate that the ASE population changes is given by the competition between the rate that dimers are formed and the rate at which they decay, or,
\begin{equation}\label{eq:ASEcycleDynamics}
\frac {\partial n_{ASE} (t)} {\partial t} = - f \left(\frac {\sigma I_P} {\hbar \omega} \right) n_{ASE}(t) + \beta n_{dimer} ,
\end{equation}
where $\sigma$ is the absorption cross section, $\hbar \omega$ is the photon energy of the pump, $I_p$ is the intensity of the pump, and $f = \gamma_{25} / \gamma_{21}$ is the fraction of the population that leaks into the dimer population per excitation event that leads to dimer formation.  $\beta$ is the rate at which the dimer population breaks apart into molecules, and $n_{dimer} = 1 - n_{ASE}$.  Expressing Equation \ref{eq:ASEcycleDynamics} in the form \begin{equation}\label{eq:ASEcycleDynamicsSimple}
\frac {\partial n_{ASE} (t)} {\partial t} = - \alpha I_P n_{ASE}(t) + \beta \left(1 - n_{ASE}(t) \right) ,
\end{equation}
the solution for the decay process in the presence of the pump with $n_{ASE}(0)=1$ is of the form
\begin{equation}\label{nDecay}
n_{ASE} = \frac {\beta} {\beta + \alpha I} + \frac {\alpha I } {\beta + \alpha I} \cdot e^{- \left( \beta + \alpha I \right) t },
\end{equation}
and the recovery process, when the pump is turned off, is given by
\begin{equation}\label{nRecover}
n_{ASE} = 1 - \left( 1-n_{ASE}(t_0) \right) e^{- \beta t },
\end{equation}
where $t_0$ is the time at which the pump laser is turned off and $n_{ASE}(t_0)$ is the population of molecules given by Equation \ref{nDecay} at that time.

The ASE intensity will be proportional to the population of excited state tautomers which is proportional to $n_{ASE}$ and the intensity of the pump.  However, the ``constant" of proportionality is a function of the ASE intensity, which is implicitly a function of the excited state tautomer population.  The exact solution of the problem thus requires that the set of equations \ref{eq:n1} to \ref{eq:n5} be solved numerically.  For the purposes of our semi-quantitative analysis, we assume that over the range of measured ASE intensities, the constant of proportionality is approximately constant so that Equations \ref{nDecay} and \ref{nRecover} hold.

Similarly, we propose that this semi-quantitative model can be applied to any complex linear- or nonlinear-optical process that leads to a population of damaged material that does not participate in the process, as shown in Figure \ref{fig:DO11-two-level}b. The underlying idea is that the details of the complex process are unimportant in building an approximate model of the long-time dynamics.  Indeed, this same model was used to describe decay and recovery of two-photon fluorescence in a dye-doped polymer.\cite{zhu07.01}

\begin{figure}[htb]
\centerline{\includegraphics{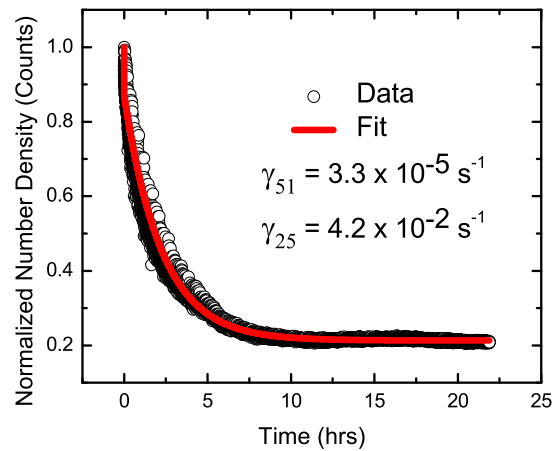}} \caption{
\label{fig:level} Photodegradation as a function of time fitted to
Equation \ref{nDecay} over time scales that are long enough to
reach equilibrium between the photodegradation rate and the
recovery rate.}
\end{figure}
Our model is characterized by the competition between an optical process that depletes the population, and an intensity-independent recovery process.  This competition leads to a state of equilibrium when the dimer population becomes large enough so that the photodegradation rate is balanced by the recovery rate.  Figure \ref{fig:level} shows the degradation of ASE in DO11 over a long-enough time scale to reach equilibrium, and a fit to Equation \ref{nDecay}.  The model and data agree within experimental uncertainties, as quantified by the scatter in the data.

\begin{figure}[htb]
\centerline{\includegraphics{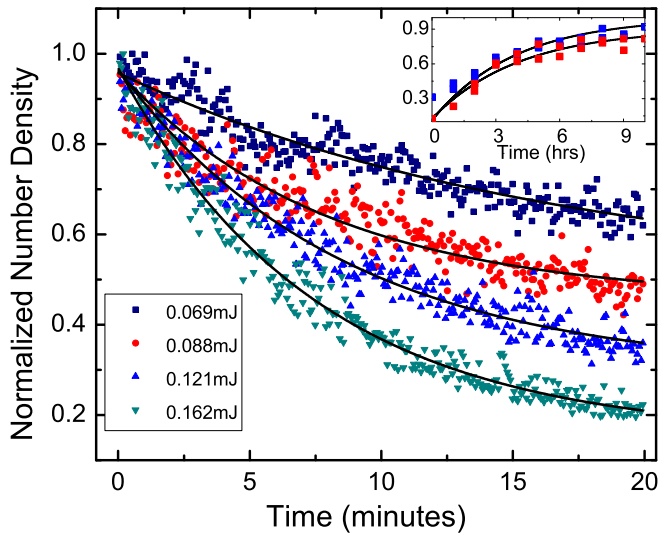}} \caption{
\label{fig:fitted_Intensity} Photodegradation as a function of
time and intensity fit to Equation \ref{eq:n3}. The inset shows
recovery results fitted to Equation\ref{eq:n5}}
\end{figure}

Photodegradation and recovery of the ASE intensity was measured as a function of energy per pulse, which is proportional to the pump intensity.  According to the model, the decay time constant increases linearly with the  pump intensity, but the recovery time is independent of the intensity history.  Figure \ref{fig:fitted_Intensity} shows the decay data along with fits to the theory.  The inset shows the recovery data.  The recovery time constants found in this data as well as from Figure \ref{fig:level} are the same within experimental uncertainties, i.e. $\gamma_{51} (\equiv \alpha) = 3.5(\pm 0.1) \times 10^{-5} s^{-1}$, as predicted with our model.

\section{Conclusion}

We have used amplified spontaneous emission and linear absorption spectroscopy during photodegradation and recovery to study the mechanisms of self healing.  The observed absorption and fluorescence spectra in DO11/PMMA show features that are suggestive of a tautomer - a species that is known to form in this class of molecules; but, can also be explained by relaxation of the excited state through dielectric interactions with the polymer.  Self-healing is observed to improve with increased DO11 concentration, and, we speculate that the tautomers can interact with each other electrostatically though dipole forces or through intermolecular proton transfer; or, combinations of the two.  As such, we have proposed that dimerization is responsible for the observed quenching of fluorescence, leading to decay in the ASE signal over time.  The dimer hypothesis is consistent with the observed splitting of the main absorption peak in the linear absorption spectrum that correlates with reduced ASE.  Alternatively, the temperature dependence that we observe is consistent with typical proton exchange energies.  The common features of all possible mechanisms is that the degradation pathway involves the formation of an aggregate.

It is plausible that a mechanism other than dimer formation may be responsible for degradation and self healing.  Whatever mechanism is responsible, the degradation process involves the formation of a population of quasi-stable ``damaged" molecules that do not fluoresce and that decay slowly into DO11 molecules over time.  We have introduced a general model that treats the process, independent of the underlying details.

Preliminary data shows that the self-healing process can be accelerated at elevated temperature, and that the ASE signal in a pristine sample drops when the temperature is elevated.  These observations are consistent with our proposed energy level diagram, as follows.  The tunneling rate from the dimer state to the molecule state should increase with temperature given that increased thermal energies increase the probability that a dimer will break up.  Similarly, a pristine sample whose population is predominantly DO11 molecules in their ground state will form dimers when the temperature is increased enough to make the thermal energy comparable to the energy difference between dimer and DO11 molecule.  Future studies will focus on the temperature dependence of both ASE and the absorption spectrum to determine the relative ground state energies of the three species.

Given that molecules in solution or in low concentrations in a polymer degrade irreversibly, our observations imply that the polymer host plays an important role in making the photodegradation process reversible.  Understanding the mechanisms of the process in different systems is an important part of building a broader theory of the more general phenomena of making irreversible processes reversible.  Furthermore, the observation that optical cycling of a material can make it more resistant to future damage may find applications in devices that need to operate near a material's damage threshold.

{\bf Acknowledgements: } We thank the National Science Foundation (ECCS-0756936) and Wright Patterson Air Force Base for generously
supporting this work.

\bibliography{\bibs}

\begin{thebibliography}{22}
\expandafter\ifx\csname natexlab\endcsname\relax\def\natexlab#1{#1}\fi
\expandafter\ifx\csname bibnamefont\endcsname\relax
  \def\bibnamefont#1{#1}\fi
\expandafter\ifx\csname bibfnamefont\endcsname\relax
  \def\bibfnamefont#1{#1}\fi
\expandafter\ifx\csname citenamefont\endcsname\relax
  \def\citenamefont#1{#1}\fi
\expandafter\ifx\csname url\endcsname\relax
  \def\url#1{\texttt{#1}}\fi
\expandafter\ifx\csname urlprefix\endcsname\relax\def\urlprefix{URL }\fi
\providecommand{\bibinfo}[2]{#2}
\providecommand{\eprint}[2][]{\url{#2}}

\bibitem[{\citenamefont{Exarhos et~al.}(1998)\citenamefont{Exarhos, Guenther,
  Kozlowski, and Soileau}}]{exarh98.01}
\bibinfo{author}{\bibfnamefont{G.~J.} \bibnamefont{Exarhos}},
  \bibinfo{author}{\bibfnamefont{A.~H.} \bibnamefont{Guenther}},
  \bibinfo{author}{\bibfnamefont{M.~R.} \bibnamefont{Kozlowski}},
  \bibnamefont{and} \bibinfo{author}{\bibfnamefont{M.~J.}
  \bibnamefont{Soileau}}, \emph{\bibinfo{title}{Laser-Induced Damage in Optical
  Materials}} (\bibinfo{publisher}{SPIE}, \bibinfo{address}{Bellevue},
  \bibinfo{year}{1998}).

\bibitem[{\citenamefont{Wood}(2003)}]{wood03.01}
\bibinfo{author}{\bibfnamefont{R.~M.} \bibnamefont{Wood}},
  \emph{\bibinfo{title}{Laser-Induced Damage of Optical Materials}}, Series in
  Optics and Optoelectronics (\bibinfo{publisher}{Taylor \& Francis},
  \bibinfo{address}{Boca Raton}, \bibinfo{year}{2003}).

\bibitem[{\citenamefont{Kuzyk}(2006)}]{kuzyk06.06}
\bibinfo{author}{\bibfnamefont{M.~G.} \bibnamefont{Kuzyk}},
  \emph{\bibinfo{title}{Polymer Fiber Optics: materials, physics, and
  applications}}, vol. \bibinfo{volume}{117} of \emph{\bibinfo{series}{Optical
  science and engineering}} (\bibinfo{publisher}{CRC Press},
  \bibinfo{address}{Boca Raton}, \bibinfo{year}{2006}).

\bibitem[{\citenamefont{Peng et~al.}(1998)\citenamefont{Peng, Xiong, and
  Chu}}]{Peng98.01}
\bibinfo{author}{\bibfnamefont{G.~D.} \bibnamefont{Peng}},
  \bibinfo{author}{\bibfnamefont{Z.}~\bibnamefont{Xiong}}, \bibnamefont{and}
  \bibinfo{author}{\bibfnamefont{P.~L.} \bibnamefont{Chu}},
  \bibinfo{journal}{J. Lightwave Technol.} \textbf{\bibinfo{volume}{16}},
  \bibinfo{pages}{2365} (\bibinfo{year}{1998}).

\bibitem[{\citenamefont{Howell and Kuzyk}(2002)}]{howel02.01}
\bibinfo{author}{\bibfnamefont{B.}~\bibnamefont{Howell}} \bibnamefont{and}
  \bibinfo{author}{\bibfnamefont{M.~G.} \bibnamefont{Kuzyk}},
  \bibinfo{journal}{J. Opt. Soc. Am. B} \textbf{\bibinfo{volume}{19}},
  \bibinfo{pages}{1790} (\bibinfo{year}{2002}).

\bibitem[{\citenamefont{Howell and Kuzyk}(2004)}]{howel04.01}
\bibinfo{author}{\bibfnamefont{B.}~\bibnamefont{Howell}} \bibnamefont{and}
  \bibinfo{author}{\bibfnamefont{M.~G.} \bibnamefont{Kuzyk}},
  \bibinfo{journal}{Appl. Phys. Lett.} \textbf{\bibinfo{volume}{85}},
  \bibinfo{pages}{1901} (\bibinfo{year}{2004}).

\bibitem[{\citenamefont{Zhu et~al.}(2007{\natexlab{a}})\citenamefont{Zhu, Zhou,
  and Kuzyk}}]{zhu07.01}
\bibinfo{author}{\bibfnamefont{Y.}~\bibnamefont{Zhu}},
  \bibinfo{author}{\bibfnamefont{J.}~\bibnamefont{Zhou}}, \bibnamefont{and}
  \bibinfo{author}{\bibfnamefont{M.~G.} \bibnamefont{Kuzyk}},
  \bibinfo{journal}{Opt. Lett.} \textbf{\bibinfo{volume}{32}},
  \bibinfo{pages}{958} (\bibinfo{year}{2007}{\natexlab{a}}).

\bibitem[{\citenamefont{Zhu et~al.}(2007{\natexlab{b}})\citenamefont{Zhu, Zhou,
  and Kuzyk}}]{zhu07.02}
\bibinfo{author}{\bibfnamefont{Y.}~\bibnamefont{Zhu}},
  \bibinfo{author}{\bibfnamefont{J.}~\bibnamefont{Zhou}}, \bibnamefont{and}
  \bibinfo{author}{\bibfnamefont{M.~G.} \bibnamefont{Kuzyk}},
  \bibinfo{journal}{Optics and Photonics News} \textbf{\bibinfo{volume}{18}},
  \bibinfo{pages}{31} (\bibinfo{year}{2007}{\natexlab{b}}).

\bibitem[{\citenamefont{Kuzyk et~al.}(2007)\citenamefont{Kuzyk, Taylor, Embaye,
  Zhe, and Zhou}}]{kuzyk07.02}
\bibinfo{author}{\bibfnamefont{M.~G.} \bibnamefont{Kuzyk}},
  \bibinfo{author}{\bibfnamefont{E.~W.} \bibnamefont{Taylor}},
  \bibinfo{author}{\bibfnamefont{N.}~\bibnamefont{Embaye}},
  \bibinfo{author}{\bibfnamefont{Y.}~\bibnamefont{Zhe}}, \bibnamefont{and}
  \bibinfo{author}{\bibfnamefont{J.}~\bibnamefont{Zhou}},
  \bibinfo{journal}{Proc. SPIE} \textbf{\bibinfo{volume}{6713}},\bibinfo{pages}{671308-1}
  (\bibinfo{year}{2007}).

\bibitem[{\citenamefont{Taylor et~al.}(2005)\citenamefont{Taylor, Nichter,
  Nash, Haas, Szep, Michalak, Flusche, Cook, McEwen, McKeon
  et~al.}}]{taylo05.01}
\bibinfo{author}{\bibfnamefont{E.~W.} \bibnamefont{Taylor}},
  \bibinfo{author}{\bibfnamefont{J.~E.} \bibnamefont{Nichter}},
  \bibinfo{author}{\bibfnamefont{F.~D.} \bibnamefont{Nash}},
  \bibinfo{author}{\bibfnamefont{F.}~\bibnamefont{Haas}},
  \bibinfo{author}{\bibfnamefont{A.~A.} \bibnamefont{Szep}},
  \bibinfo{author}{\bibfnamefont{R.~J.} \bibnamefont{Michalak}},
  \bibinfo{author}{\bibfnamefont{B.~M.} \bibnamefont{Flusche}},
  \bibinfo{author}{\bibfnamefont{P.~R.} \bibnamefont{Cook}},
  \bibinfo{author}{\bibfnamefont{T.~A.} \bibnamefont{McEwen}},
  \bibinfo{author}{\bibfnamefont{B.~F.} \bibnamefont{McKeon}},
  \bibinfo{author}{\bibfnamefont{P.~M.} \bibnamefont{Payson}},
  \bibinfo{author}{\bibfnamefont{G.~A.} \bibnamefont{Brost}}, and
  \bibinfo{author}{\bibfnamefont{A.~R.} \bibnamefont{Pirich}},
  \bibinfo{journal}{Appl. Phys. Lett.}
  \textbf{\bibinfo{volume}{86}}, \bibinfo{pages}{201122}
  (\bibinfo{year}{2005}).

\bibitem[{\citenamefont{Smith et~al.}(1991)\citenamefont{Smith, Zaklika,
  Thakur, Walker, Tominaga, and Barbara}}]{smith91.01}
\bibinfo{author}{\bibfnamefont{T.~P.} \bibnamefont{Smith}},
  \bibinfo{author}{\bibfnamefont{K.~A.} \bibnamefont{Zaklika}},
  \bibinfo{author}{\bibfnamefont{K.}~\bibnamefont{Thakur}},
  \bibinfo{author}{\bibfnamefont{G.~C.} \bibnamefont{Walker}},
  \bibinfo{author}{\bibfnamefont{K.}~\bibnamefont{Tominaga}}, \bibnamefont{and}
  \bibinfo{author}{\bibfnamefont{P.~F.} \bibnamefont{Barbara}},
  \bibinfo{journal}{J. Phys. Chem.} \textbf{\bibinfo{volume}{95}},
  \bibinfo{pages}{10465} (\bibinfo{year}{1991}).

\bibitem[{\citenamefont{Shynkar et~al.}(2003)\citenamefont{Shynkar, Mely,
  Duportail, Piemont, Klymchenko, and Demchenko}}]{shynk03.01}
\bibinfo{author}{\bibfnamefont{V.~V.} \bibnamefont{Shynkar}},
  \bibinfo{author}{\bibfnamefont{Y.}~\bibnamefont{Mely}},
  \bibinfo{author}{\bibfnamefont{G.}~\bibnamefont{Duportail}},
  \bibinfo{author}{\bibfnamefont{E.}~\bibnamefont{Piemont}},
  \bibinfo{author}{\bibfnamefont{A.~S.} \bibnamefont{Klymchenko}},
  \bibnamefont{and} \bibinfo{author}{\bibfnamefont{A.~P.}
  \bibnamefont{Demchenko}}, \bibinfo{journal}{J. Phys. Chem.}
  \textbf{\bibinfo{volume}{107}}, \bibinfo{pages}{9522} (\bibinfo{year}{2003}).

\bibitem[{\citenamefont{Ghebremichael et~al.}(1993)\citenamefont{Ghebremichael,
  Kuzyk, and Dirk}}]{ghebr93.01}
\bibinfo{author}{\bibfnamefont{F.}~\bibnamefont{Ghebremichael}},
  \bibinfo{author}{\bibfnamefont{M.~G.} \bibnamefont{Kuzyk}}, \bibnamefont{and}
  \bibinfo{author}{\bibfnamefont{C.~W.} \bibnamefont{Dirk}},
  \bibinfo{journal}{Nonlinear Optics} \textbf{\bibinfo{volume}{6}},
  \bibinfo{pages}{123} (\bibinfo{year}{1993}).

\bibitem[{\citenamefont{Ghebremichael and Kuzyk}(1995)}]{ghebr95.04}
\bibinfo{author}{\bibfnamefont{F.}~\bibnamefont{Ghebremichael}}
  \bibnamefont{and} \bibinfo{author}{\bibfnamefont{M.~G.} \bibnamefont{Kuzyk}},
  \bibinfo{journal}{J. Appl. Phys.} \textbf{\bibinfo{volume}{77}},
  \bibinfo{pages}{2896} (\bibinfo{year}{1995}).

\bibitem[{\citenamefont{Poga and Kuzyk}(1994)}]{poga94.01}
\bibinfo{author}{\bibfnamefont{C.}~\bibnamefont{Poga}} \bibnamefont{and}
  \bibinfo{author}{\bibfnamefont{M.~G.} \bibnamefont{Kuzyk}},
  \bibinfo{journal}{J. Opt. Soc. Am. B} \textbf{\bibinfo{volume}{11}},
  \bibinfo{pages}{80} (\bibinfo{year}{1994}).

\bibitem[{\citenamefont{Poga et~al.}(1995)\citenamefont{Poga, Brown, Kuyzk, and
  Dirk}}]{poga95.01}
\bibinfo{author}{\bibfnamefont{C.}~\bibnamefont{Poga}},
  \bibinfo{author}{\bibfnamefont{T.~M.} \bibnamefont{Brown}},
  \bibinfo{author}{\bibfnamefont{M.~G.} \bibnamefont{Kuyzk}}, \bibnamefont{and}
  \bibinfo{author}{\bibfnamefont{C.~W.} \bibnamefont{Dirk}},
  \bibinfo{journal}{J. Opt. Soc. Am. B} \textbf{\bibinfo{volume}{12}},
  \bibinfo{pages}{531} (\bibinfo{year}{1995}).

\bibitem[{\citenamefont{Garvey et~al.}(1996{\natexlab{a}})\citenamefont{Garvey,
  Li, Kuzyk, Dirk, and Martinez}}]{garve96.02}
\bibinfo{author}{\bibfnamefont{D.~W.} \bibnamefont{Garvey}},
  \bibinfo{author}{\bibfnamefont{Q.}~\bibnamefont{Li}},
  \bibinfo{author}{\bibfnamefont{M.~G.} \bibnamefont{Kuzyk}},
  \bibinfo{author}{\bibfnamefont{C.~W.} \bibnamefont{Dirk}}, \bibnamefont{and}
  \bibinfo{author}{\bibfnamefont{S.}~\bibnamefont{Martinez}},
  \bibinfo{journal}{Opt. Lett.} \textbf{\bibinfo{volume}{21}},
  \bibinfo{pages}{104} (\bibinfo{year}{1996}{\natexlab{a}}).

\bibitem[{\citenamefont{Garvey et~al.}(1996{\natexlab{b}})\citenamefont{Garvey,
  Zimmerman, Young, Tostenrude, Townsend, Zhou, Lobel, Dayton, Wittorf, and
  Kuzyk}}]{garve96.01}
\bibinfo{author}{\bibfnamefont{D.~W.} \bibnamefont{Garvey}},
  \bibinfo{author}{\bibfnamefont{K.}~\bibnamefont{Zimmerman}},
  \bibinfo{author}{\bibfnamefont{P.}~\bibnamefont{Young}},
  \bibinfo{author}{\bibfnamefont{J.}~\bibnamefont{Tostenrude}},
  \bibinfo{author}{\bibfnamefont{J.~S.} \bibnamefont{Townsend}},
  \bibinfo{author}{\bibfnamefont{Z.}~\bibnamefont{Zhou}},
  \bibinfo{author}{\bibfnamefont{M.}~\bibnamefont{Lobel}},
  \bibinfo{author}{\bibfnamefont{M.}~\bibnamefont{Dayton}},
  \bibinfo{author}{\bibfnamefont{R.}~\bibnamefont{Wittorf}}, \bibnamefont{and}
  \bibinfo{author}{\bibfnamefont{M.~G.} \bibnamefont{Kuzyk}},
  \bibinfo{journal}{J. Opt. Soc. Am. B} \textbf{\bibinfo{volume}{13}},
  \bibinfo{pages}{2017} (\bibinfo{year}{1996}{\natexlab{b}}).

\bibitem[{\citenamefont{El-Bayoumi et~al.}(1975)\citenamefont{El-Bayoumi,
  Avouris, and Ware}}]{ashra75.01}
\bibinfo{author}{\bibfnamefont{M.~A.} \bibnamefont{El-Bayoumi}},
  \bibinfo{author}{\bibfnamefont{P.}~\bibnamefont{Avouris}}, \bibnamefont{and}
  \bibinfo{author}{\bibfnamefont{W.~R.} \bibnamefont{Ware}},
  \bibinfo{journal}{J. Chem. Phys.} \textbf{\bibinfo{volume}{62}},
  \bibinfo{pages}{2499} (\bibinfo{year}{1975}).

\bibitem[{\citenamefont{Ingham and El-Bayoumi}(1973)}]{ingha73.01}
\bibinfo{author}{\bibfnamefont{K.~C.} \bibnamefont{Ingham}} \bibnamefont{and}
  \bibinfo{author}{\bibfnamefont{M.~A.} \bibnamefont{El-Bayoumi}},
  \bibinfo{journal}{J. Am. Chem. Soc.} \textbf{\bibinfo{volume}{96}},
  \bibinfo{pages}{1674} (\bibinfo{year}{1974}).

\bibitem[{\citenamefont{Embaye}(2007)}]{embay07.01}
\bibinfo{author}{\bibfnamefont{N.}~\bibnamefont{Embaye}}, Ph.D. thesis,
  \bibinfo{school}{Washington State University} (\bibinfo{year}{2007}).

\bibitem[{\citenamefont{Ernsting and Nikolaus}(1986)}]{ernsr86.01}
\bibinfo{author}{\bibfnamefont{N.~P.} \bibnamefont{Ernsting}} \bibnamefont{and}
  \bibinfo{author}{\bibfnamefont{B.}~\bibnamefont{Nikolaus}},
  \bibinfo{journal}{Appl. Phys. B} \textbf{\bibinfo{volume}{39}},
  \bibinfo{pages}{155} (\bibinfo{year}{1986}).

\end{thebibliography}

\end{document}